\documentclass[a4paper,11pt]{article}
\usepackage{pos}

\title{The fundamental limit of jet tagging: Beyond top jets}
\author[a]{Sarah Koller}
\author*[a]{Humberto Reyes-González}

\affiliation[a]{Institute for Theoretical Particle Physics and Cosmology, RWTH Aachen University, Aachen, Germany}

\emailAdd{humberto.reyes@rwth-aachen.de}

\abstract{
Jet tagging, i.e.\ determining the origin of high-energy hadronic jets, is a key challenge in particle physics. Machine-learning-based taggers have achieved remarkable progress, raising the question of how close current methods are to the theoretical limit of performance. Ref.~\cite{Geuskens:2024tfo} addressed this question for boosted top-quark jets using transformer-based generative models that provide realistic synthetic jet data with known probability density functions. This enables a direct comparison between modern taggers and the optimal likelihood-ratio classifier. In this note, we summarize the approach and extend the study to boosted W, Z, and H$\rightarrow gg$ jets. We find that the gap to the estimated optimal limit is strongly jet dependent and is substantially reduced for these seemingly more challenging tagging tasks. Ongoing work aimed at understanding the interpretation, robustness, and scaling of these limits is also briefly discussed.
}

\FullConference{23rd International Workshop on Advanced Computing and Analysis Techniques in Physics Research (ACAT2025)\\
8--12 September 2025\\
Hamburg, Germany\\}

\begin{document}
\maketitle

\section{Introduction}

Machine-learning-based jet tagging has become a central component of the LHC physics program, enabling the identification of heavy particles through their hadronic decay products. Over the past decade, increasingly sophisticated architectures have led to substantial improvements in performance. This naturally raises a fundamental question: how close are current taggers to the best possible classifier allowed by the information contained in the jet?

From the Neyman--Pearson lemma, the optimal discriminator between two jet classes $A$ and $B$ is given by the likelihood ratio $p(x|A)/p(x|B)$, or equivalently by the log-likelihood ratio $\log p(x|A)-\log p(x|B)$, where $x$ denotes a representation of the jet. Any classifier can at best approximate this quantity. In practice, however, the underlying probability densities are not known analytically. Ref.~\cite{Geuskens:2024tfo} proposed a framework to address this problem using autoregressive transformers with explicit density estimation. Trained on simulated jet samples, these models provide a generative approximation to the underlying probability density functions and enable the construction of synthetic datasets with known likelihoods. This allows direct comparisons between machine-learning classifiers and the optimal likelihood-ratio test.

In this note, we summarize the method and extend the study of boosted top jets ($t\rightarrow bqq'$) to boosted $W\rightarrow qq'$, $Z\rightarrow q\bar{q}$, and $H\rightarrow gg$ jets. These additional benchmarks allow us to compare how the achievable performance and the gap to the estimated optimum depend on the underlying jet structure. We also briefly discuss the interpretation and robustness of these limits, together with ongoing work aimed at understanding their origin.

\section{The setup}

Following Ref.~\cite{Finke:2023veq}, autoregressive transformers are trained on the JetClass dataset~\cite{Qu:2022mxj}, which contains approximately 10 million training events and 2 million testing events for each of ten jet classes. Here we focus on top, $W$, $Z$, and $H$ jets, using QCD jets as background.

The models are trained at constituent level using $(p_T,\eta,\phi)$ information. After discretization into $(40,30,30)$ bins, each constituent is represented by a token, resulting in a vocabulary of roughly 39k tokens. A separate transformer is trained for each jet class, providing an explicit approximation to its probability density function.

After training, 10 million synthetic events are generated per class. Since the models now provide exact likelihood values for the generated events, the optimal discriminator can be constructed from
\begin{equation}
    \log \frac{p_S(x)}{p_{\mathrm{QCD}}(x)},
\end{equation}
where $p_S$ and $p_{\mathrm{QCD}}$ denote the densities estimated by the signal and QCD models, respectively. The log likelihood-ratio test (LLR) defines the optimal ROC curve. The same synthetic datasets are then used to train the baseline transformer classifier (BTC), which shares the same backbone architecture as the generative model but is equipped with a classification head. This enables a direct comparison with the estimated theoretical optimum under controlled conditions.

\section{Results}

\begin{figure}[h]
    \centering
    \begin{minipage}{0.56\linewidth}
        \centering
        \includegraphics[width=\linewidth]{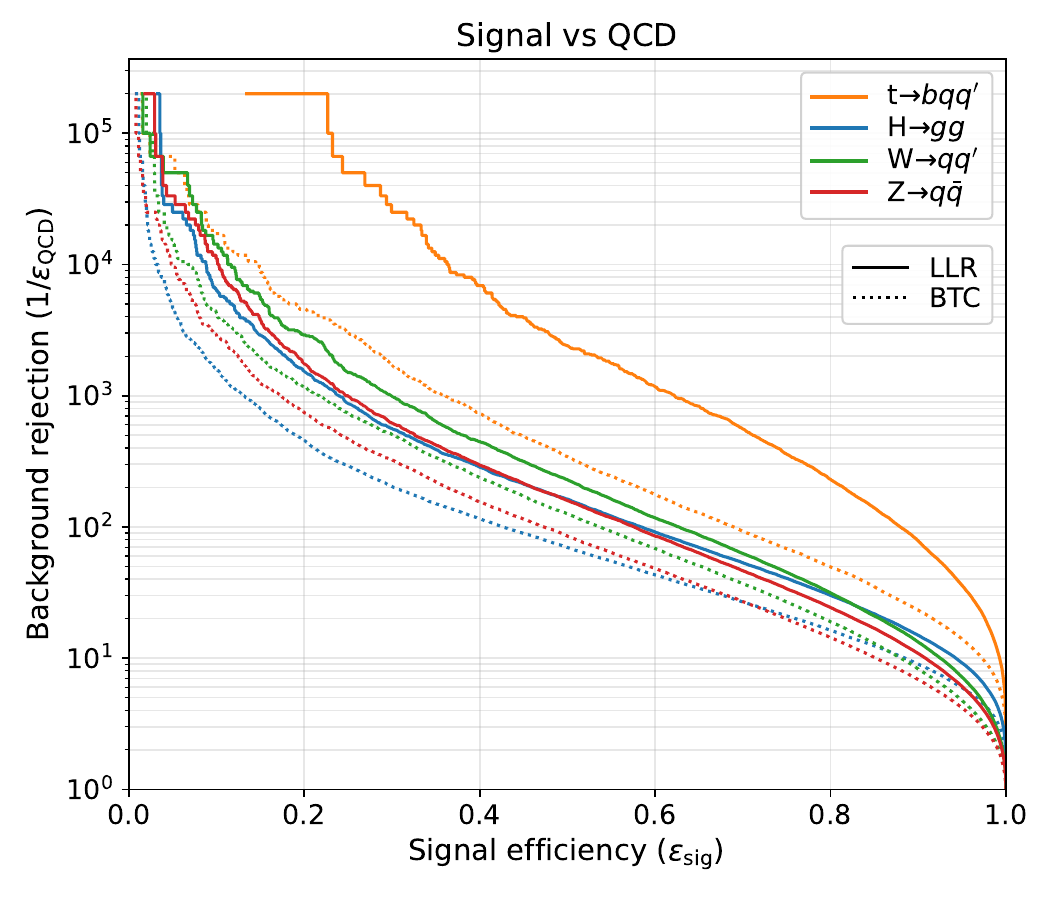}
        \caption{ROC curves comparing the optimal classifier (solid) and the baseline transformer classifier (dotted) for top, W, Z, and H$\rightarrow gg$ jet tagging.}
        \label{fig:alljets}
    \end{minipage}
    \hfill
    \begin{minipage}{0.40\linewidth}
        \centering
        \includegraphics[width=\linewidth]{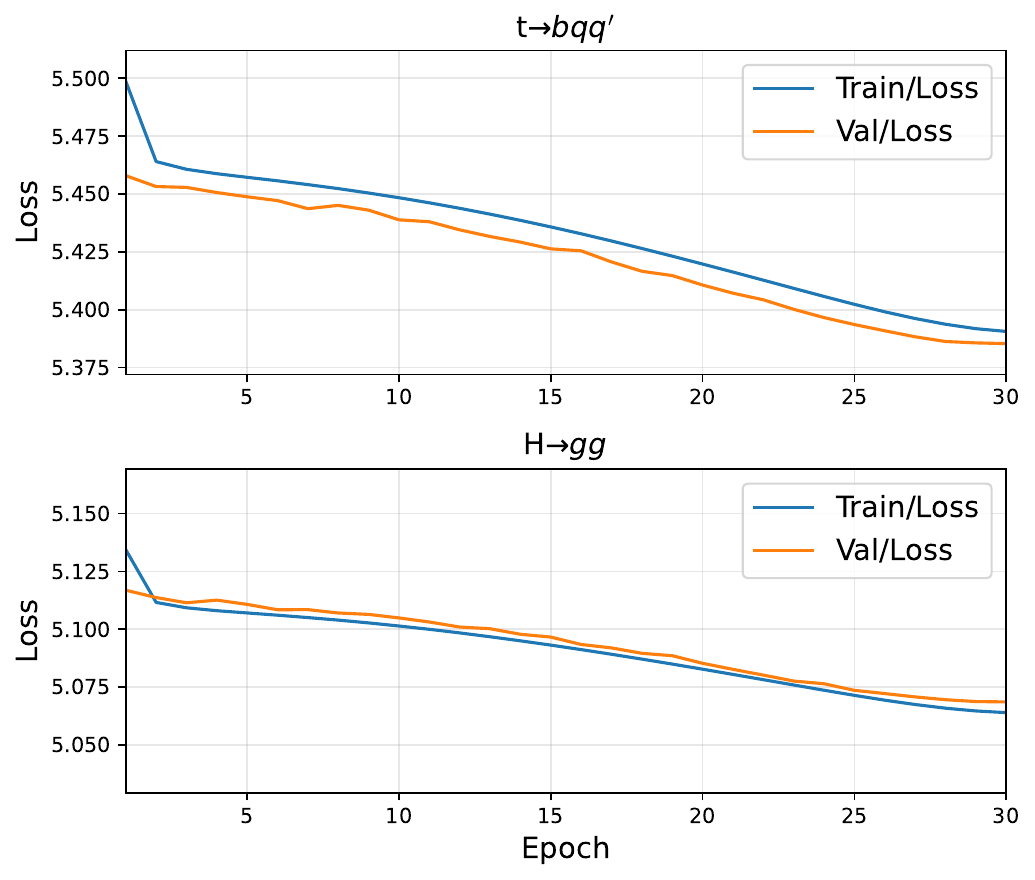}
        \caption{Representative training and validation loss curves for the generative models trained on top (upper) and $H\rightarrow g g$ (lower) jets. No evidence of conventional overfitting is observed.}
        \label{fig:loss_curves}
    \end{minipage}
\end{figure}

Figure~\ref{fig:alljets} compares the optimal classifier with the BTC for boosted top, $W$, $Z$, and $H$ jet tagging. The sizeable gap previously observed for top tagging~\cite{Geuskens:2024tfo} remains visible. In contrast, the gap is substantially smaller for $W$, $Z$, and $H\rightarrow gg$ jets and is nearly closed in some cases. This is particularly interesting because these tagging tasks are often regarded as more challenging in terms of their absolute discrimination performance.

A possible interpretation is that $W$, $Z$, and $H\rightarrow gg$ jets exhibit less distinctive substructure and are closer to QCD jets, both morphologically and in the stochasticity of their radiation patterns. Although this reduces the overall discrimination power, it also limits the amount of exploitable information, making the optimal limit easier to approach. Top jets, by contrast, exhibit richer features, including a characteristic three-prong topology and the presence of a $b$ quark, leaving more room for improvement beyond current classifiers.

Representative training histories are shown in Fig.~\ref{fig:loss_curves}. The displayed models show no evidence of conventional overfitting. Nevertheless, the interpretation of the estimated optimal limit requires care. Ref.~\cite{Pang:2025lbs} highlighted that the inferred limit can depend on the choice and quality of the generative model, and additional nuisances may affect its interpretation. At this stage, our main aim is therefore not to assign a definitive value to the optimal limit, but to compare different jet classes within a common framework and highlight the strongly reduced gap for W, Z, and $H\rightarrow gg$ jets.

\section{Outlook}\label{outlook}

The framework of Ref.~\cite{Geuskens:2024tfo} provides a practical way to estimate the fundamental limit of jet tagging and quantify the remaining gap to optimal performance. Extending the study beyond top jets reveals that this gap is strongly process dependent and can be nearly closed for $W\rightarrow qq'$, $Z\rightarrow q \bar{q}$, and $H\rightarrow gg$ jets.

Current work focuses on understanding how the estimated limit evolves with increasing training statistics and improved optimization of the underlying generative models, as well as identifying robust convergence criteria for the learned probability densities. We are also investigating systematic validation of these densities using statistical hypothesis testing techniques~\cite{Grossi:2024axb,Grossi:2025pmm,Cappelli:2025myc}. Connections to recent studies of scaling laws in jet generation~\cite{Amram:2026zzv} and classification~\cite{Vigl:2026ppx} are also under investigation. Finally, we aim to better understand the origin of the larger gap observed for top tagging and explore whether it can be quantitatively related to the intrinsic stochasticity of different jet classes.

\acknowledgments

This work was supported by the SciFM consortium (05D25PA5) funded by the German Federal Ministry of Research, Technology, and Space (BMFTR) in the ErUM-Data action plan, as well as by the German Research Foundation (DFG) under grant 396021762 -- TRR 257: Particle Physics Phenomenology after the Higgs Discovery.

%%%%%%%%%%%%%%%%%%%%%%%%%%%%%%%%%%%%%%%%%%%%%%%%%%%%%%%%%%%%%%%%%%%%%%%%%%%%%%%

% --- BIBLIOGRAPHY SETTINGS ---
\bibliographystyle{JHEP}
\bibliography{references}
% ------------------------------

\end{document}